\documentstyle[12pt]{article}

\def\al{\alpha}

\def\be{\begin{equation}}

\def\ee{\end{equation}}

\def\bea{\begin{eqnarray}}

\def\eea{\end{eqnarray}}

\def\kt{k_\perp}
\def\kta{k_{a\perp}}
\def\ktb{k_{b\perp}}

\begin{document}

\topskip 2cm 
\begin{titlepage}

\hspace*{\fill}\parbox[t]{4cm}{EDINBURGH 96/7\\ June 1996}

\vspace{2cm}

\begin{center}
{\large\bf Forward Jets at HERA and at the Tevatron} \\
\vspace{1.5cm}
{\large Vittorio Del Duca} \\
\vspace{.5cm}
{\sl Particle Physics Theory Group,\,
Dept. of Physics and Astronomy\\ University of Edinburgh,\,
Edinburgh EH9 3JZ, Scotland, UK}\\
\vspace{1.5cm}
\vfil
\begin{abstract}
\noindent
In this talk I consider forward-jet production
at HERA and at the Tevatron as a probe of the multiple
gluon radiation induced by the BFKL evolution.
\end{abstract}

\vspace{2cm}

{\sl To appear in the Proceedings of\\ the International Workshop on\\
Deep Inelastic Scattering and Related Phenomena\\
Roma, Italy, April 1996}
\end{center}

\end{titlepage}

\baselineskip=0.8cm

\section{Introduction}
\label{sec:uno}

In DIS at HERA semi-hard processes, for which the squared
center-of-mass energy $s$ is much larger than the momentum transfer
$Q^2$, are investigated and values of $x_{bj}=Q^2/s$ of the
order of $10^{-5}$ have been attained \cite{hera}. The evolution of the 
$F_2(x_{bj},Q^2)$ structure function in $\ln(Q^2)$ is usually described 
by the DGLAP equation. However, at very small values of $x_{bj}$ we may 
consider to resum the leading logarithmic (LL) contributions in $1/x_{bj}$
to $F_2$, by using the BFKL evolution equation \cite{bal,bcm}. 

A caveat is in order: the BFKL equation computes the
radiative corrections to parton-parton scattering in the high-energy limit,
assuming that the outgoing partons are balanced in transverse momentum.
Therefore there is no transverse-momentum evolution in the process. 
It is not possible to assess whether this costraint is fulfilled in the 
configurations that drive the rise of $F_2$ at small $x_{bj}$, but it may be 
forced upon the DIS process by tagging a jet in the 
proton direction \cite{muel} and by requiring that the squared jet 
transverse momentum is 
of the order of $Q^2$. An analogous process for which the multiple
gluon radiation induced by the BFKL evolution may be relevant 
is dijet production at large rapidity intervals 
$\eta$ in $p\,\bar p$ collisions \cite{mn}.

\section{Forward jets}
\label{sec:due}

In the high-energy limit parton-parton scattering
is dominated by gluon exchange in the cross channel, which is typically an
${\cal O}(\al_s^2)$ process. This occurs at leading order (LO) in 
dijet production in $p\,\bar p$ collisions~\cite{EKS}, and at 
next-to-leading order (NLO) in dijet production with a forward jet in 
DIS~\cite{mz} (or at LO in three-jet production~\cite{mz,us}). Indeed, 
in DIS the NLO dijet production and the LO three-jet production
turn out to be bigger~\cite{mz} than the LO dijet production, which is 
${\cal O}(\al_s)$, when a forward jet is required.
On top of the cross-channel-gluon dominated processes, the BFKL equation 
resums the LL contributions, in $\ln(\hat s/Q^2)$, to all orders
in $\al_s$ in the multi-Regge kinematics, which assumes that the outgoing
partons are strongly ordered in 
rapidity $\eta$ and have comparable transverse momentum.
The higher-order corrections to gluon exchange yield a gluon ladder
in the cross channel \cite{bal}. The leading logarithms are resummed 
by the function,
\be
f(\kta,\ktb,\tilde{\phi},\eta)\, =\, {1\over (2\pi)^2 \kta 
\ktb} \sum_{n=-\infty}^{\infty} 
e^{in\tilde{\phi}}\, \int_{-\infty}^{\infty} d\nu\, 
e^{\omega(\nu,n)\eta}\, \left(\kta^2\over\ktb^2\right)^{i\nu}\, ,\label{solc}
\ee
with $\bf\kta$ and $\bf\ktb$ the transverse momenta of the gluons
at the ends of the ladder, $\tilde\phi$ the azimuthal angle
between them, $\eta\simeq\ln(\hat s/\kt^2)$ an evolution parameter of
the ladder required to be large, and $\omega(\nu,n)$ the eigenvalue of
the BFKL equation whose maximum $\omega(0,0)=4\ln{2}N_c\al_s/\pi$
yields the known power-like growth of $f$ in energy~\cite{bal}.

In inclusive dijet production in $p\,\bar p$ collisions
the resummed parton cross section is~\cite{mn,DS},
\be
{d\hat \sigma\over d\kta^2 d\ktb^2 d\phi} = {\pi N_c^2 \al_s^2 \over 2
\kta^2 \ktb^2}\, f(\kta^2,\ktb^2,\tilde{\phi},\eta)\, ,\label{ppbar}
\ee
with $\phi$ the azimuthal angle between the tagging jets, 
$\phi=\tilde{\phi}+\pi$. At the hadron level, $\eta$ is the rapidity 
difference between the tagging jets, $\eta=\eta_{j_1}-\eta_{j_2}$, and 
accordingly evidence of the BFKL dynamics is searched in dijet events at large
rapidity intervals~\cite{d0}.

In forward-jet production in DIS in the lab frame the lepton-parton
cross section is~\cite{muel,us,me}
\be
{d\hat \sigma\over dy dQ^2 d\kt^2 d\phi} = \sum_q e_q^2 {N_c\al^2 
\al_s^2 \over \pi^2 (Q^2)^2 \kt^2 y}
\int {dv_{\perp}^2\over v_{\perp}^2} f(v_{\perp}^2,\kt^2,\tilde{\phi},\eta) 
{\cal F}(v_{\perp}^2,Q^2,\hat{\phi},y)\, ,\label{hot}
\ee
with $y$ the electron energy loss; $Q^2$ the photon vituality; the function
${\cal F}$ accounting for the $q\,\bar q$ pair that in the high-energy limit
mediates the scattering between the photon and the cross-channel gluon; 
$\kt$ and $v_{\perp}$ respectively the transverse momenta of the 
forward jet and of the gluon attaching to the $q\,\bar q$ pair; 
$\hat{\phi}$ the azimuthal angle between the
photon and the gluon; $\phi$ the azimuthal angle between the outgoing 
electron and the jet, with $\phi=\hat{\phi}+\tilde{\phi}+\pi$; 
and with the sum over the quark flavors in the $q\,\bar q$ pair.
$\eta$ is then related to $x_{bj}$ and to the momentum fraction $x$ of the
parton initiating the hard scattering within the proton through 
$\eta=\ln(x/x_{bj})$. Producing the jet forward
ensures that $x$ is not small; $\eta$ is then
made large by selecting events at small $x_{bj}$.

The BFKL ladder $f$ (\ref{solc}) induces a strong enhancement in the parton 
cross sections (\ref{ppbar}) and (\ref{hot}) when $\eta$ grows~\cite{bal}. 
In a hadron collider $\eta=\eta_{j_1}-\eta_{j_2}\simeq\ln(x_1x_2s/\kt^2)$. 
At fixed $s$, like at the Tevatron, $\eta$ grows by increasing $x_1$ and 
$x_2$. This introduces a damping in the production rate, due to the 
falling parton luminosity \cite{DS}, and conceals the growth due to
$f$ (\ref{solc}). The advantage of HERA is that a
fixed-energy $ep$ collider is nonetheless a variable-energy collider 
in the photon-proton frame \cite{muel}, thus it is possible to increase 
$\eta=\ln(x/x_{bj})$ by decreasing $x_{bj}$ while keeping fixed $x$.

The truncation to ${\cal O}(\al_s^2)$ of the forward-jet rate derived from
eq.~(\ref{hot}), which has three final-state partons, corresponds to the 
lowest-order approximation to the BFKL ladder and is in 
good agreement with the exact LO three-jet rate with a forward jet~\cite{us}. 
However, the BFKL calculation
with the full ladder (\ref{solc}) yields a curve whose normalization is
bigger by an order of magnitude, and which grows faster than the
${\cal O}(\al_s^2)$ evaluations as $x_{bj}$
decreases. The H1-Collaboration data~\cite{H1} seem to favor the BFKL
calculation~\cite{us}. This looks encouraging, however as a caveat we recall
that in dijet production at the Tevatron
a comparison of the ${\cal O}(\al_s^3)$ matrix elements, exact and
in the BFKL approximation, shows that the latter overestimates the
available phase space~\cite{DDS}.

\section{The azimuthal-angle decorrelation}
\label{sec:tre}

In two-jet production at large rapidity intervals in $p\,\bar p$ collisions
the BFKL evolution predicts that the $\phi$ correlation between the tagging 
jets decreases as the rapidity difference between the tagging jets 
increases~\cite{DS}. This phenomenon has been
observed by the D0 Collaboration at the Tevatron~\cite{d0}, 
however, the BFKL ladder yields too much decorrelation
while the Monte Carlo JETRAD~\cite{ggk}, based on the exact NLO dijet 
production, yields too little decorrelation.
The data is in good agreement with a simulation from the Monte 
Carlo HERWIG~\cite{march}. This seems to suggest that
corrections higher than ${\cal O}(\al_s^3)$ are needed to describe the
data, but not so much of it as contained in the BFKL ladder.

In jet production in DIS
we know that at the parton-model level, i.e. at $x=x_{bj}$, the jet and
the electron are produced back-to-back, and we expect that when $x > x_{bj}$,
but with $\eta=\ln(x/x_{bj})$ still small, the jet production is dominated at
the parton level by the photon-gluon fusion diagram, which has two final-state 
partons and is expected to yield the usual correlation at $\phi=\pi$
between the electron and the parton tagged as the jet. 
However as $\eta$ grows the jet production is increasingly
dominated by diagrams with three-final state partons and with gluon
exchange in the cross channel, and eventually by the higher-order 
corrections to them induced by the BFKL ladder.

The lowest-order approximation to the BFKL ladder yields a $\phi$ distribution 
peaked at $\phi=\pi/2$. Implementing
then the full BFKL ladder (\ref{solc}), the $\phi$ correlation is
completely washed out~\cite{us}. However, the high-energy limit can not
appreciate the transition from a distribution peaked at $\phi=\pi$ at
large $x_{bj}$ to one peaked at $\phi=\pi/2$ at small $x_{bj}$. It
was suggested that an exact ${\cal O}(\al_s^2)$ calculation should see 
it~\cite{me}, and indeed it does~\cite{us}. It is to be seen if it will 
also be observed in the data.

\section*{Acknowledgments}
I thank PPARC and the Travel and Research Committee of the
University of Edinburgh for the support, and the organizers of DIS96
for the hospitality.



\begin{thebibliography}{99}

\bibitem{hera} H1 Collab., {\em Nucl.~Phys.}~B {\bf 407}, 515 (1993); 
{\em Nucl.~Phys.}~B {\bf 439}, 471 (1995); preprint DESY 96-039;\\
ZEUS Collab., {\em Phys.~Lett.}~B {\bf 316}, 412 (1993); {\em Zeit.~Phys.}~C 
{\bf 65}, 369 (1995); {\em Zeit.~Phys.}~C {\bf 69}, 607 (1996).

\bibitem{bal} E.A.~Kuraev, L.N.~Lipatov and V.S.~Fadin, {\em 
Zh.~Eksp.~Teor.~Fiz.} {\bf 72}, 377 (1977) [{\em Sov.~Phys.~JETP} {\bf 45}, 
199 (1977)]; Ya.Ya.~Balitsky and L.N.~Lipatov, {\em Yad.~Fiz.} {\bf 28} 
1597 (1978) [{\em Sov.~J.~Nucl.~Phys.} {\bf 28}, 822 (1978)].

\bibitem{bcm} A.~Bassetto, M.~Ciafaloni and G.~Marchesini, {\em Phys.~Rep.}
{\bf 100}, 201 (1983).

\bibitem{muel}A.H.~Mueller, {\em Nucl.Phys.}~B (Proc.Suppl.) {\bf 18}C, 125
(1991);\\ W.-K.~Tang, {\em Phys.~Lett.}~B {\bf 278}, 363 (1991);\\
J.~Bartels, A.~De~Roeck and M.~Loewe, {\em Z.~Phys.}~C {\bf 54}, 635 (1992);\\
J.~Kwiecinski, A.D.~Martin and P.J.~Sutton, {\em Phys.~Rev.}~D {\bf 46}, 921
(1992).

\bibitem{mn}A.H.~Mueller and H.~Navelet, {\em Nucl.~Phys.}~B {\bf 282}, 
727 (1987).

\bibitem{EKS}S.D.~Ellis, Z.~Kunszt and D.E.~Soper, {\em Phys.~Rev.~Lett.} 
{\bf 69}, 1496 (1992). 

\bibitem{mz}E.~Mirkes and D.~Zeppenfeld, preprint TTP96-10.

\bibitem{us}J.~Bartels, V.~Del~Duca, A.~De~Roeck, D.~Graudenz and 
M.~W\"usthoff, preprint DESY-96-36.

\bibitem{DS}V.~Del~Duca and C.R.~Schmidt, {\em Phys.~Rev.}~D {\bf 49}, 
4510 (1994); {\em Nucl.~Phys.}~B (Proc. Suppl.) {\bf 39}C, 137 (1995);
\\ W.J.~Stirling, {\em Nucl.~Phys.}~B {\bf 423}, 56 (1994).

\bibitem{d0}D0 Collab., preprint FERMILAB-PUB-96/038-E.

\bibitem{me}V.~Del~Duca, Proc. of the Workshop ``Deep
Inelastic Scattering and QCD'', Paris, France, 1995, Editions de l'Ecole
Polytechnique 1995.

\bibitem{H1}H1 Collab., {\em Phys.~Lett.}~B {\bf 356}, 118 (1995).

\bibitem{DDS}V.~Del~Duca and C.R.~Schmidt, {\em Phys.~Rev.}~D {\bf 51}, 2150
(1995).

\bibitem{ggk}W.T.~Giele, E.W.N.~Glover and D.A.~Kosower, {\em Nucl.~Phys.}~B 
{\bf 403}, 633 (1993); {\em Phys.~Rev.~Lett.} {\bf 73}, 2019 (1994).

\bibitem{march}G.~Marchesini and B.R.~Webber, {\em Nucl.~Phys.}~B {\bf 310},
461 (1988);\\ G.~Marchesini {\it et al.}, {\em Comp.~Phys.~Comm.} {\bf 67},
465 (1992).

\end{thebibliography}
\end{document}